\documentclass[onecolumn,a4,showpacs,preprintnumbers,amsmath,amssymb]{revtex4-1}
\usepackage{mathtools}
\usepackage{graphicx}
\usepackage{dcolumn}
\usepackage{bm}
\usepackage{url}
\begin{document}
\def\1qpc{$1{\rm-}qpc$}
\def\2qpc{$2{\rm-}qpc$}
\def\3qpc{$3{\rm-}qpc$}
\title{
Longitudinal Wobbling in $^{133}$La
}


\author{S. Biswas,$^{1}$ R. Palit,$^{1}$ S. Frauendorf,$^{2}$ U. Garg,$^{2}$ W. Li 
,$^{2}$ G. H. Bhat,$^{3}$   J. A. Sheikh,$^{3}$ 
J. Sethi,$^{1}$ S. Saha,$^{1}$ Purnima Singh,$^{1}$ D. Choudhury,$^{1}$ J. T. Matta,$^{2}$ A. D. Ayangeakaa,$^{2}$
W. A. Dar,$^{3}$ V. Singh,$^{4}$ S. Sihotra$^{4}$}
	
\address{
$^{1}$Department of Nuclear and Atomic Physics, Tata Institute of Fundamental Research, Mumbai 400005, India\\
$^{2}$University of Notre Dame, Indiana 46556, USA\\
$^{3}$Department of Physics, University of Kashmir, Srinagar 190006, Indiana\\
$^{4}$Department of Physics, Panjab University, Chandigarh 160014, India}

\date{\today}    

\begin{abstract}

Excited states of $^{133}$La have been investigated to search for the wobbling
excitation mode in the low-spin regime. Wobbling bands with $n_\omega$ = 0
and 1 are identified along with the interconnecting $\Delta I$ = 1, $E2$ 
transitions, which are regarded as one of the characteristic features of the wobbling motion. An 
increase in wobbling frequency 
with spin implies longitudinal wobbling for $^{133}$La, in contrast with the
case of transverse wobbling observed in  $^{135}$Pr. This is the first observation 
of a longitudinal wobbling band in nuclei. The experimental 
observations are accounted for by calculations using the 
quasiparticle-triaxial-rotor (QTR) model,
which attribute the appearance of longitudinal wobbling
to the early alignment of a $\pi=+$ proton pair.
\end{abstract}

\maketitle

The atomic nucleus is a fascinating mesoscopic system,
which continues to reveal new collective excitation modes due to
improved sensitivity in the experimental techniques~\cite{cl97, je02, iw14}. 
For quadrupole deformed nuclei, loss of axial symmetry generates new types of collective excitations.
The novel characteristic rotational features of triaxial nuclei are chirality
and wobbling. 
The focus of the present communication is the wobbling mode, which emerges
because a triaxial nucleus can carry collective angular momentum along all three principal axes.
 In contrast, an axial nucleus  cannot rotate about its symmetry axis.
The wobbling mode is well known for triaxial-rotor molecules~\cite{Herzberg}. It appears as a family  of rotational bands 
 based on excitations  with increasing angular momentum components along the  two axes with smaller moments of inertia. 
Its appearance in even-even triaxial nuclei was analyzed by Bohr and Mottelson~\cite{bohr}. 
It is characterized by an excitation energy, $E(I)~(=\hbar\omega)$ increasing with the angular momentum $I$  and enhanced collective $\Delta I$ = 1, $E2$
interconnecting transitions with reduced transition probabilities proportional to 1/$I$. 

The evidence for this 
mode in nuclei is rather limited. In a survey~\cite{mo06} across the nuclear chart using the finite-range liquid-drop model 
(FRLDM), soft triaxial ground-state shapes are predicted for nuclides around $Z$ = 62, $N$ = 76; $Z$ = 44, $N$ = 64; and $Z$ = 78, $N$ = 116.
As reviewed in Ref.~\cite{SF15}, most of these nuclides seem to execute large-amplitude oscillations only around the axial shape.
Partial evidence for stabilization of ground state triaxiality 
in these regions has been the observation of odd-$I$-low
staggering of the quasi-$\gamma$ band~\cite{st07, zh09, lu13, to13}. 
Ref.~\cite{fr14} discussed the best case, $^{112}$Ru: 
the quasi-$\gamma$ band  splits at high spin into the one- and two- phonon wobbling bands with the expected 
$\omega \propto I$. However, the data in $^{112}$Ru~\cite{zh09} do not contain the required information about the 
$B(E2, I\rightarrow I-1)$ values to clearly establish the wobbling character.

More compelling evidence for 
the wobbling phenomenon was observed   in the $A\sim$160 mass region at high spin in the  odd-
$A $ Lu and Ta isotopes~\cite{od01, sc03, am03, br05, ha09} and at low spin in the $A \sim$130
mass region in $^{135}$Pr~\cite{ma15}.
The presence of the odd quasiparticle in these nuclei modifies the wobbling mode in a substantial way,
which provides additional evidence for the triaxial shape.   Frauendorf and D\"{o}nau have recently 
analyzed the modifications semiclassically~\cite{fr14}. They distinguish between two different kinds of wobbling modes for the odd-$A$ nuclei.
For the \textquotedblleft longitudinal\textquotedblright~mode, the angular momentum of the odd particle is parallel to
the axis with the largest MoI, whereas for the \textquotedblleft transverse\textquotedblright~mode,
it is perpendicular to this axis. The two modes can be recognized by the $I$-dependence of the wobbling 
energy $E_{wob}(I)$ (= $\hbar \omega_{wob}$), which increases or decreases, respectively. It is given by 
\begin{equation}
\begin{split}
  E_{wob}(I) &= E(I, n_\omega =1)\\
  &-[E(I-1,n_\omega=0)+E(I+1,n_\omega=0)]/2.
\end{split}                
\end{equation}
The wobbling bands observed in the Lu and Ta isotopes are examples of transverse
wobbling because the wobbling frequency decreases with
increasing spin, which also holds  for the very recently reported  first observation of transverse wobbling at low spin in $^{135}$Pr \cite{ma15}.
The lowering of wobbling frequency enhances the detection probability and makes it possible  to detect
pattern of the enhanced reduced transition probabilities for interband $E2$ transitions, which is a crucial  signal for wobbling. 
In all cases the high-j quasiproton has particle character. Its interaction with the triaxial core aligns  its angular momentum
with the short axis, and the medium axis has the maximal MoI, which results in {\it transverse} wobbling (see Ref.~\cite{fr14}).   

This Letter reports on observation of {\it longitudinal} wobbling   
in the nucleus $^{133}$La, which is an isotone of $^{135}$Pr. 
The $n_\omega=1$ phonon band with band head at $13/2^-$, is found to decay to the $n_\omega=0$ 
phonon band by $\Delta I = 1$, $E2$ transitions whose multipolarities have been 
determined on the basis of directional correlation from oriented states (DCO), polarization, and angular distribution measurements. The wobbling frequency is shown 
to increase with increasing spin indicating longitudinal wobbling. 
This is the first time that a longitudinal wobbling band has been
observed in nuclei.
Using the
quasiparticle triaxial rotor (QTR) model with and without harmonic frozen approximation (HFA),
the transition from transverse to longitudinal wobbling mode  is demonstrated to be caused by the early alignment of a pair of 
positive-parity quasiprotons with the short axis. 

\begin{figure}[h]
\includegraphics[angle=270, width=9.5cm, trim = 1cm 4cm 0cm 3cm]{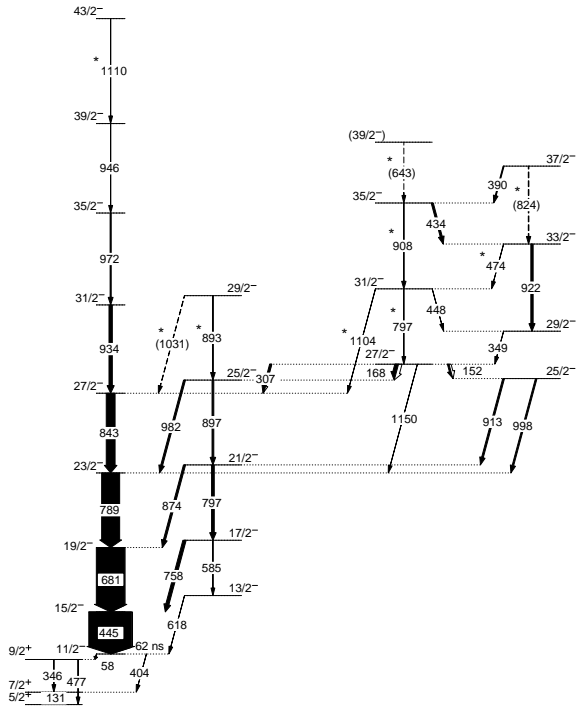}
\caption{\label{fig:1}Partial level scheme of $^{133}$La showing the previously known yrast band ($n_\omega=0$), the $n_\omega=1$ wobbling band, 
and the dipole band. The transitions marked with an asterisk are new. The intensities have been obtained from the 445-keV gated
coincidence spectrum. The intensities of the transitions are proportional to the widths of the arrows.}
\end{figure}

\begin{figure*}[ht]
\centering
\includegraphics[width=10.0cm, trim = 0cm 0cm 0cm 0cm]{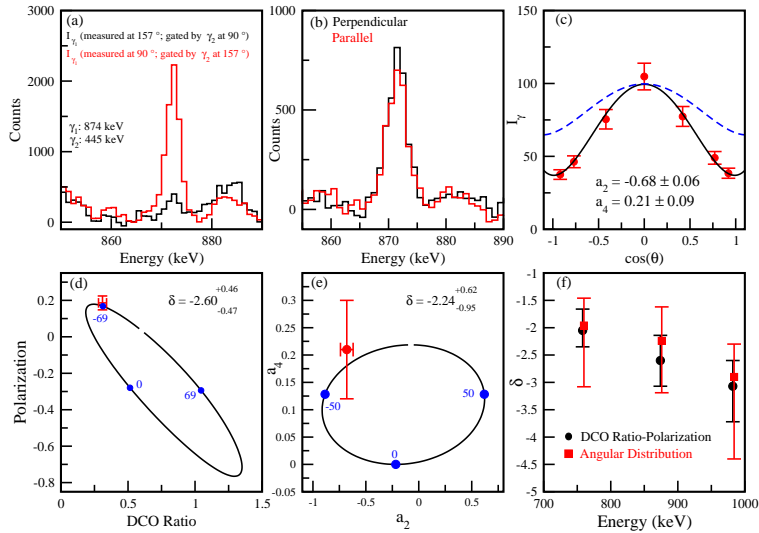}
\caption{\label{fig:2}(color online). (a) The spectra for the DCO Ratio (b) The perpendicular and parallel spectra
for polarization (c) The angular distribution plot (the red points are
the experimental points, the black solid curve shows the best fit to the data and the blue dashed curve shows the 
curve for a pure $\Delta$I = 1, M1 transition) for the 874-keV transition with a gate on 445-keV transition.
(d) Contour plot for DCO Ratio vs. Polarization as a function of the mixing ratio ($\delta$)
(the red point on the ellipse is the experimental data point and the blue points represent the $\delta$ values)
for 874-keV transition with a gate on 445-keV transition. (e) a$_{2}$-a$_{4}$ contour plot for the 874-keV transition and 
(f) Comparison of mixing ratios extracted from DCO Ratio-Polarization method (black) and Angular Distribution Method (red) for
758, 874 and 982 -keV transitions.}
\end{figure*}

A 52-MeV $^{11}$B beam from the 14-UD Pelletron at Tata Institute of Fundamental Research (TIFR) was used to populate 
the excited states of $^{133}$La via $^{126}$Te($^{11}$B, 4n) reaction. 
A layer of 1.1 mg/cm$^2$-thick enriched $^{126}$Te evaporated on
an Au backing of 9.9 mg/cm$^2$ served as the target. The emitted $\gamma$ rays were detected in Indian 
National Gamma Array (INGA), which consisted of 21 Compton suppressed clover 
HPGe detectors coupled with a digital data acquisition system~\cite{pa12}.
A set of 3$\times$10$^{8}$ two- and higher-fold events were collected during
the experiment. 
The time stamped data were sorted in a $\gamma$-$\gamma$-$\gamma$ 
cube and angle dependent $\gamma$-$\gamma$ matrices, 
and the RADWARE software package~\cite{ra95} was used for further analysis of these matrices and 
cubes. 

A partial level scheme of $^{133}$La containing the negative parity states
studied in the present work, is shown in Fig.~\ref{fig:1}. This level scheme is based on  detailed
analysis of the $\gamma$-$\gamma$-$\gamma$ coincidence relations, cross-over
transitions, and relative intensities of the concerned $\gamma$ rays. 
Spin and parity assignments to the states have been made on the basis of the 
measured DCO ratios ($R_{DCO}$) and polarization asymmetries of the transitions 
depopulating these states. 
The detectors at 90$^{\circ}$ and 157$^{\circ}$ were used to determine the DCO ratios~\cite{kr89}.
The polarization of $\gamma$ rays was extracted from the 90$^{\circ}$ detectors using the formula given in 
Refs.~\cite{st99, pa00}.

Prior to the present work, the nucleus $^{133}$La was studied through heavy-ion 
fusion evaporation reactions using a small detector array~\cite{hi91}.
The present work confirms the previous results. In addition, we have observed 
several new $\gamma$-ray transitions; these have been marked with an asterisk in Fig.~\ref{fig:1}. The yrast and the yrare bands have been 
extended to $I^\pi = 43/2^-$ and $I^\pi = 29/2^-$ and identified as the 
$n_\omega$ = 0 and $n_\omega$ = 1 phonon wobbling bands, respectively.
A dipole band has also been observed at high spin up to $I^\pi = 39/2^-$. There are a few transitions, not
shown in Fig.~\ref{fig:1}, that were reported in previous work \cite{hi91} as feeding into the 
$n_\omega$ =1 band.
While it would be tempting to speculate that those might be from a possible $n_\omega$ = 2 band, they are 
too weak to obtain any conclusive evidence regarding their multipolarities, based on angular distributions and polarization asymmetries.


\begin{table*}
\caption{\label{tab:table1}The mixing ratios ($\delta$), $E2$ fractions (=$\delta^2$/(1+$\delta^2$))
and the experimental and theoretical transition probability ratios $\frac {B(M1_{out})}{B(E2_{in})}$ and $\frac{B(E2_{out})}{B(E2_{in})}$
for the transitions from the $n_\omega$ = 1 to the $n_\omega$ = 0 band in $^{133}$La.
Experimental ratios have been obtained from measured intensities of the transitions.}
 \centering
 \scalebox{0.8}{
 \begin{tabular}{c c c c c c c c c c c}
  \\\hline\hline

  I$_{i}^{\pi}\rightarrow$ I$_{f}^{\pi}$ & E$_{\gamma}$ & $\delta$(Expt.) & $\delta$ (QTR) &$E2$ Fraction (\%)& $\frac
  {B(M1_{out})}{B(E2_{in})}$ (Expt.)& $\frac
  {B(M1_{out})}{B(E2_{in})}$ (QTR)& $\frac{B(E2_{out})}{B(E2_{in})}$ (Expt.)&
$\frac{B(E2_{out})}{B(E2_{in})}$ (QTR$/$HFA)        \\
  \hline
  13/2$^{-}\rightarrow$ 11/2$^{-}$ & 618 & -1.48$^{+0.45}_{-0.32}$&-0.67 & 68.6$^{+13.1}_{-9.3}$ & -                                        &   0.665    & -                                          &  1.158/0.299        \\\\
  17/2$^{-}\rightarrow$ 15/2$^{-}$ & 758 & -2.05$^{+0.39}_{-0.30}$ &-0.94 &80.8$^{+5.9}_{-4.5}$ & 0.107$^{+0.035}_{-0.028}$&  0.358   & 1.127$^{+0.140}_{-0.130}$&0.774  /0.324          \\\\
  21/2$^{-}\rightarrow$ 19/2$^{-}$ & 874 & -2.60$^{+0.46}_{-0.47}$ &--1.20 &87.1$^{+4.0}_{-4.0}$ & 0.056$^{+0.018}_{-0.019}$&  0.231     & 0.716$^{+0.079}_{-0.079}$&  0.591 /0.311      \\\\
  25/2$^{-}\rightarrow$ 23/2$^{-}$ & 982 & -3.07$^{+0.47}_{-0.65}$ &-1.52 &90.4$^{+2.6}_{-3.7}$ & 0.039$^{+0.011}_{-0.015}$ &   0.162   &0.545$^{+0.057}_{-0.059}$&   0.496 / 0.269     \\\\
  29/2$^{-}\rightarrow$ 27/2$^{-}$ & 1031 & - & -1.92& - & -                                                                                                             &   0.119    & -                                        & 0.445  /0.221               
\\
  \hline\hline
 \end{tabular}}
\end{table*}

One characteristic feature of wobbling bands is that the 
transitions between the $n_{\omega}$ = 1 and $n_{\omega}$ = 0 bands must be of $\Delta I$ = 1, $E2$ character~\cite{od01}.
The spectra which have been used to extract the DCO Ratio and Polarization for the 874-keV transition are shown in Figs.~\ref{fig:2}(a) and (b) 
respectively. Fig.~\ref{fig:2}(c) shows the angular distribution plot for this transition.
The $R_{DCO}$ and the polarization asymmetry for the 874-keV ($21/2^- \rightarrow 19/2^-$) transition
were calculated as a function of mixing ratio ($\delta$), which is the ratio of the reduced matrix elements for the $E2$ and $M1$ components
of a parity non-changing $\Delta I$ = 1 transition.
The contour plot of theoretical $R_{DCO}$ and polarization along
with the experimental values is shown in Fig.~\ref{fig:2}(d).  
In the calculation of $R_{DCO}$, the 
width of the sub-state population ($\sigma/I$) was assumed to be 0.3.
This comparison gives a mixing ratio of $-69^\circ$ for the 874-keV transition and firmly confirms its 
$\Delta I~=~1$, $E2$ nature. In  Fig.~\ref{fig:2}(e), the $a_{2}-a_{4}$ contour plot along with the experimental data point has been shown.
The mixing ratios obtained from both the DCO Ratio-Polarization and angular distribution method have been shown in Fig.~\ref{fig:2}(f) for
the 758-, 874- and 982-keV connecting transitions..
However, the mixing ratios for the linking transitions mentioned in Table~\ref{tab:table1} are from the DCO-Polarization method. 
The $E2$ fraction increases for the connecting transitions with 
increasing spin, indicating enhancement of wobbling with increasing angular momentum.

\begin{figure}[t]
\includegraphics[width=\columnwidth]{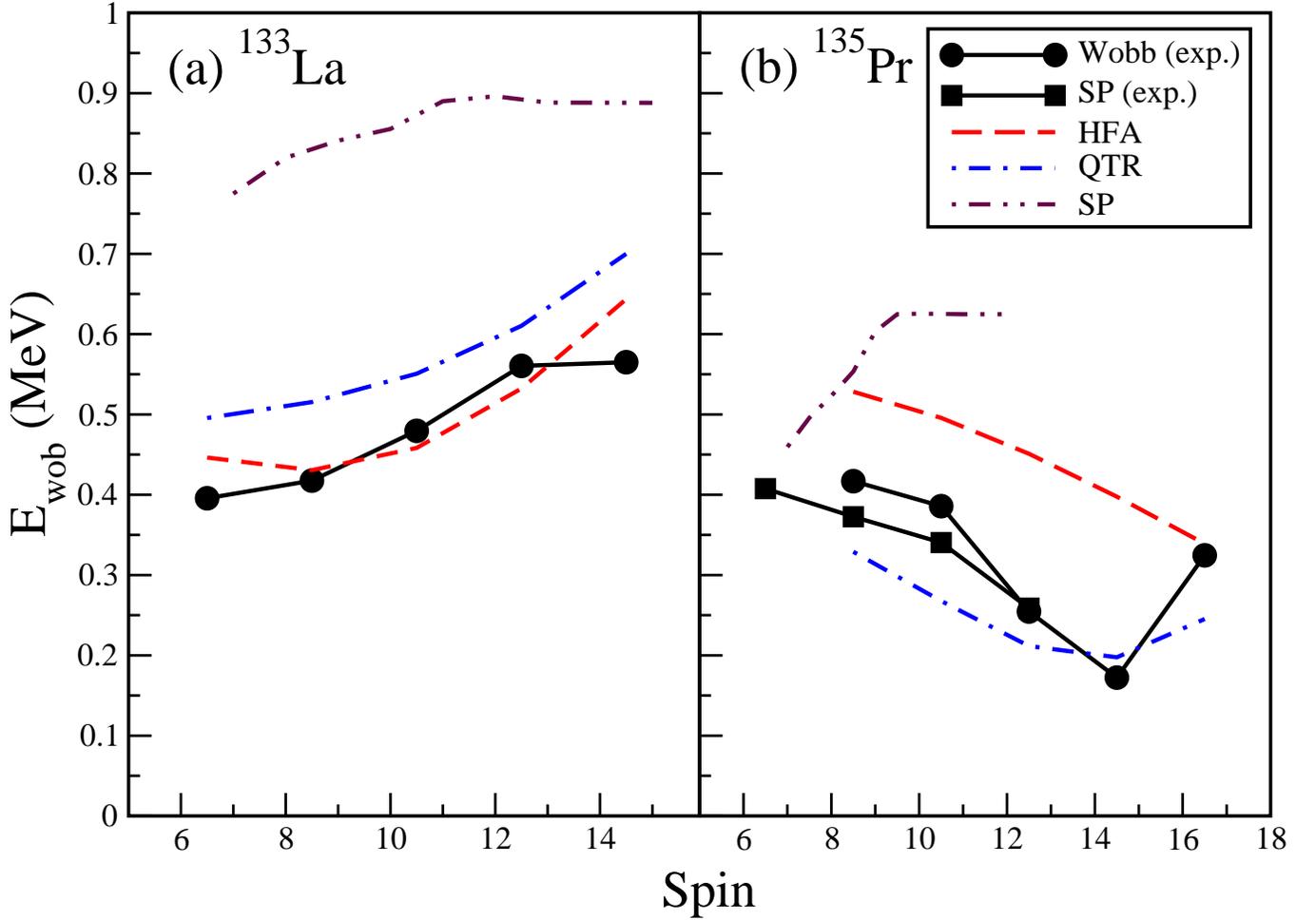}
\caption{\label{fig:3}(color online). Comparison between experiment (solid lines) and theory (HFA (dashed lines), QTR (dash-dotted lines))
of the variation of wobbling frequency for $n_\omega$ = 1 band in (a) $^{133}$La and (b) $^{135}$Pr.
The dash-dot-dot line denoted by SP shows the position of the unfavored proton h$_{11/2}$ signature partner relative to the 
favored yrast sequence, as calculated by the TAC model.}
\end{figure}

The experimental wobbling energies, $E_{wob}$ (see Eq.(1)) of the isotones $^{133}$La and $^{135}$Pr
are plotted in Fig.~\ref{fig:3} as a function of spin. 
The wobbling frequency for $^{133}$La is {\it increasing} with angular momentum, which along with
the mixing ratios of the interconnecting 
transitions suggests longitudinal wobbling in $^{133}$La~\cite{fr14}, while it is {\it decreasing}  in $^{135}$Pr, 
for which transverse wobbling has been established~\cite {ma15}.
The change from longitudinal to transverse in the neighboring isotones is a
surprising result  and warrants a detailed investigation. For both isotopes, the $h_{11/2}$ quasiproton has particle character, i. e., it
is expected to align with the short axis, which is confirmed by cranking calculations~\cite{hi91,sem86}. According to the general arguments 
of Ref. \cite{fr14}, {\it  both isotones should have been transverse}.

\begin{figure}[t]
\includegraphics[width=\columnwidth]{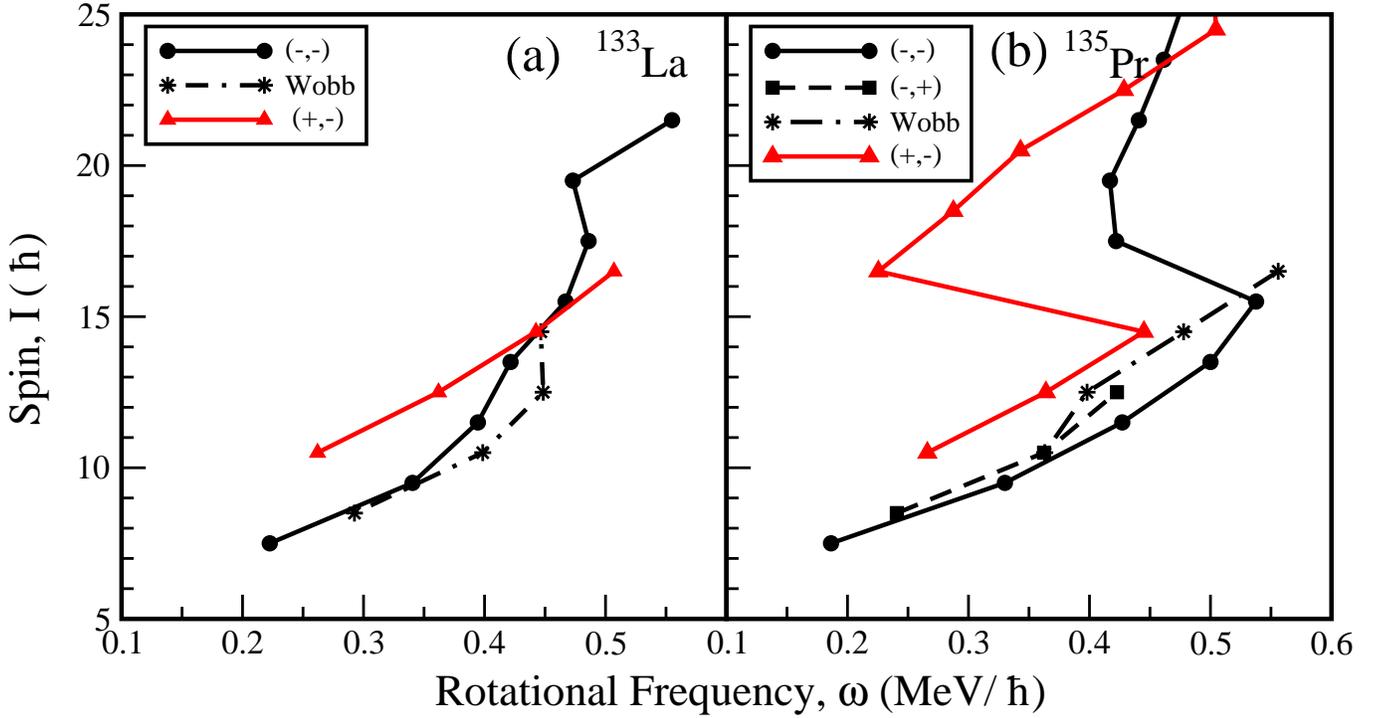}
\caption{\label{fig:4}(color online). Spin (I) {\it vs.} experimental rotational frequency ($\omega$) for 
(a) $^{133}$La and (b) $^{135}$Pr.  Shown in Black  are the h$_{11/2}$ favored signature sequence (full circle),  its unfavored 
signature partner (dash with square), and the wobbling band (dash-dot with star). The full Red curve (triangles) shows the  
positive-parity sequence $(\pi,\alpha)=(+,-1/2)$ shifted up by 5.5.}
\end{figure}

To identify the origin of this abrupt change from transverse to longitudinal wobbling motion, plots of the spin of the yrast band as function of 
the rotational frequency,
$I(\omega)$, for the two nuclei are displayed  in Fig.~\ref{fig:4}. Both curves have an $\omega=0$ intercept of about 5.5, which reflects the 
alignment of the  $h_{11/2}$
proton with the short axis. The angular momentum increases early and in a gradual manner in $^{133}$La compared to $^{135}$Pr, which shows a late
and rapid backbend.
The change of the wobbling mode is, then, understood as follows. Just above  the band head, both nuclides rotate
about the short axis to which the $h_{11/2}$ proton is aligned. For $^{135}$Pr, the MoI of 
the short axis is smaller than the MoI of the medium axis ($\mathcal J_{s}<\mathcal J_{m}$). With increasing spin it
becomes favorable to put more and more collective angular momentum on the medium
axis, which has the larger MoI. This leads to the decrease of the wobbling frequency 
with spin, the hallmark of transverse wobbling. As seen in Fig. 4 (a),  the MoI of the short axis is larger for $^{133}$La, such that the MoI of the two axes are 
about  the same ($\mathcal J_{s}\sim\mathcal J_{m}$). The medium axis is no longer preferred by the collective angular momentum,
which is now added to the short axis. This results in the increasing wobbling frequency
with spin, which is the hallmark of longitudinal wobbling.

\begin{figure}[t]
\includegraphics[width=\columnwidth]{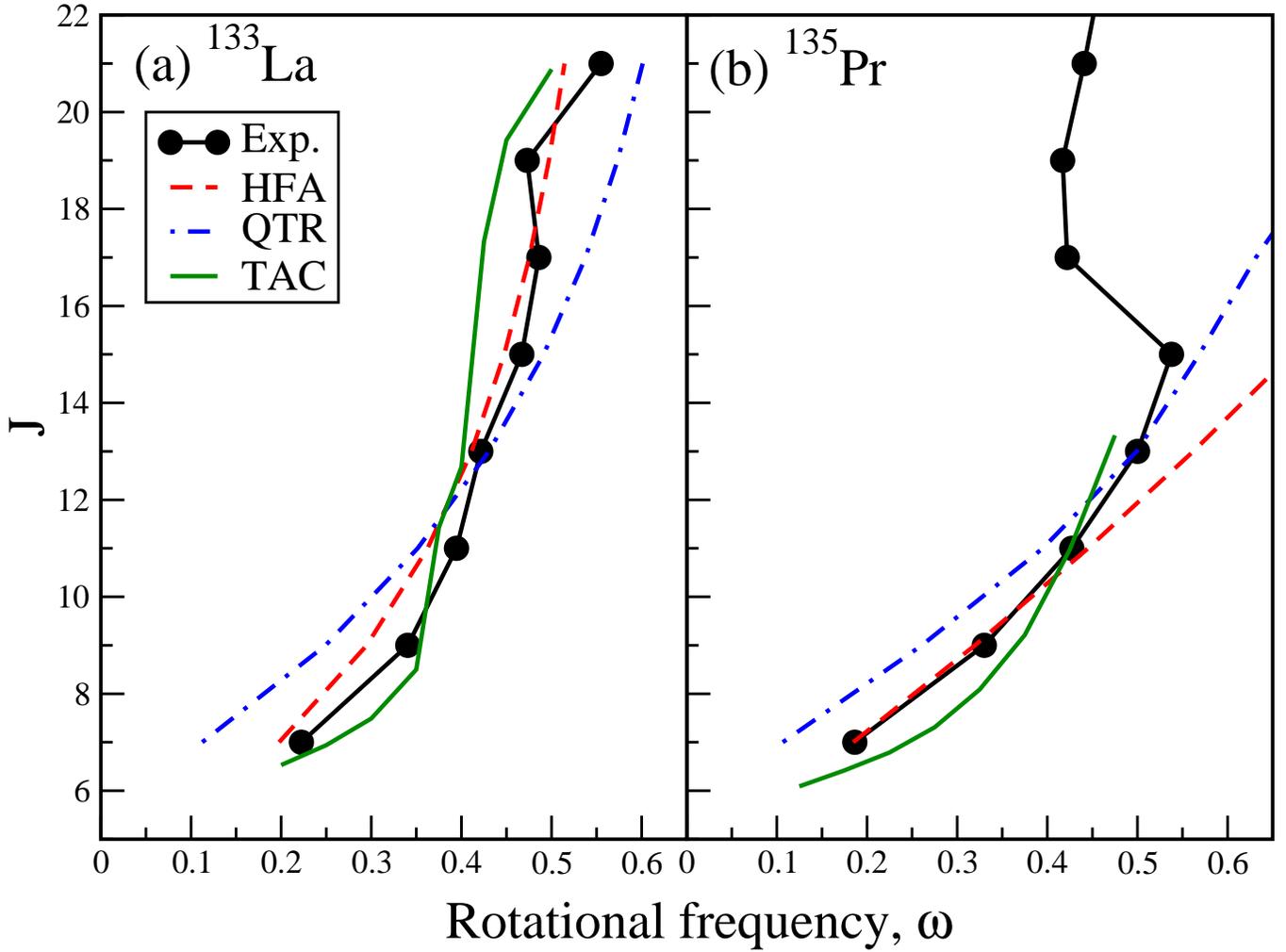}
\caption{\label{fig:5}(color online). (J = I -1/2) {\it vs.} experimental (Black circle) rotational frequency ($\omega$) for  the h$_{11/2}$ 
favored signature sequence in (a) $^{133}$La and (b) $^{135}$Pr compared with HFA (dash Red), QTR (dash-dot Blue), and TAC (full Green) calculations. }
\end{figure}

To quantify the claim,  we modified the simple QTR+HFA model of Ref.~\cite{fr14} by introducing
 a spin-dependent MoI for the short axis.  We used the expression $\mathcal J_{s}=\Theta_0+\Theta_1 R,~~~R=I-i $, where $\mathcal J_{s}$ is MoI of 
 the short axis, $I$ is the total angular momentum,
 $i$ the odd proton angular momentum, and $R$ the core angular momentum. The experimental  alignments shown in Fig.~\ref{fig:5} are well accounted
 for with the  parameters 
 $\Theta_0= 10,13 ~\hbar^2/MeV,~\Theta_1= 0.8,0.2~ \hbar/MeV$, and $i=4.5,5~\hbar$  for La and Pr, respectively.  
 The other two MoI's were fixed at $\mathcal J_{m}= 21~\hbar^2/MeV$ ($\mathcal J_{m}$ is MoI of the medium axis), 
 and $\mathcal J_{l} = 4~\hbar^2/MeV$ ($\mathcal J_{l}$ is MoI of the long axis) for both nuclides, as used in Ref. \cite {fr14}. 
 As seen in Fig.~\ref{fig:3}, the QTR+HFA calculation (depicted by HFA in the figure) reproduces the change from
 transverse wobbling in $^{135}$Pr to longitudinal in $^{133}$La.
 The reason is the early increase of the MoI in  $^{133}$La compared to  $^{135}$Pr seen in Fig.~\ref{fig:4}.  
The ratios $\mathcal J_{m}/\mathcal J_{s}/\mathcal J_{l}$  for $^{135}$Pr and $^{133}$La at $I=33/2$ are, respectively, 21/15.4/4 and 21/19/4, which correspond to
transverse and longitudinal wobbling. The QTR+HFA values for $B(E2_{out})/B(E2_{in})$ also account for the data reasonably well (see Tab. \ref{tab:table1}). 

To further refine the theoretical interpretation we modified the QTR model used  in Ref.~\cite{fr14} in the same way as described 
for its HFA approximation.   The core moments of inertia  for $^{133}$La  were $\mathcal J_{s}=\Theta_0+\Theta_1 R$ with
 $\Theta_0= 12.5\times 0.73~\hbar^2/MeV,~\Theta_1= 0.9\times 0.73~ \hbar/MeV$,
 $\mathcal J_{m}= 21\times 0.73~\hbar^2/MeV$, and $\mathcal J_{l} = 4\times 0.73~\hbar^2/MeV$.
 For $^{135}$Pr they were $\mathcal J_{s}=\Theta_0+\Theta_1 R$ with
 $\Theta_0= 12.8~\hbar^2/MeV,~\Theta_1= 0.14~ \hbar/MeV$,
 $\mathcal J_{m}= 21~\hbar^2/MeV$, and $\mathcal J_{l} = 4~\hbar^2/MeV$. A core-particle coupling strength of 6.4 was used, which 
 corresponds to a deformation of $\varepsilon=0.16$ and $\gamma=26^{\circ}$, which is the equilibrium deformation found in the TAC calculations
 and used for the triaxial rotor core. 

The QTR also accounts for 
the change from transverse to longitudinal wobbling (see Fig. \ref{fig:3}) and the difference in alignment between the nuclei (see Fig. \ref{fig:5}).
The QTR values for $B(M1)_{out}/B(E2)_{in}$, $B(E2)_{out}/B(E2)_{in}$,
 and the mixing ratios $\delta$, also account for the data reasonably well (see Table \ref{tab:table1}).  
The decrease of the ratio $B(M1)_{out}/B(E2)_{in}$ is reproduced as well; however QTR overestimates the ratio by factor of three.
 Importantly,  the collective enhancement of the 
non-stretched $E2$ transitions from the wobbling to yrast band  is born out by the QTR.
The $E2$ fraction indicates that the transitions are $E2$ dominated.  The negative sign of $\delta$
is consistent with the orientation of the $h_{11/2}$ quasiproton along the short axis.

We attribute the early increase of angular momentum in $^{133}$La to the gradual alignment of a pair of positive-parity quasiprotons 
of (dg) nature with the short axis, which is at variance with  previous interpretations ~\cite{hi91,sem86}. Fig. \ref{fig:4} shows that the positive-parity band 
based on the odd (dg) quasiproton does not show the early rise of angular momentum seen in the $h_{11/2}$ yrast band, because it is blocked by the odd proton. 
The curve displayed in the figure is shifted by 
5.5 $\hbar$, the amount contributed by the aligned $h_{11/2}$. It crosses the yrast sequence about half-way the up-bend, which is expected when one 
quasiproton of the gradually aligning pair is excited. In case of $^{135}$Pr, the shifted curve of the positive parity (dg) keeps a relative distance of about
3$\hbar$ to the yrast sequence
until it runs into a backbend, caused by the alignment of other quasiparticles. The fact that its continuation crosses the yrast sequence may suggest that the
alignment of a pair of 
(dg) quasiprotons participates in the backbend.  

\begin{figure}[h]
\vspace{2cm}
\includegraphics[width=8.5cm, trim = 0cm 0cm 0cm 3cm]{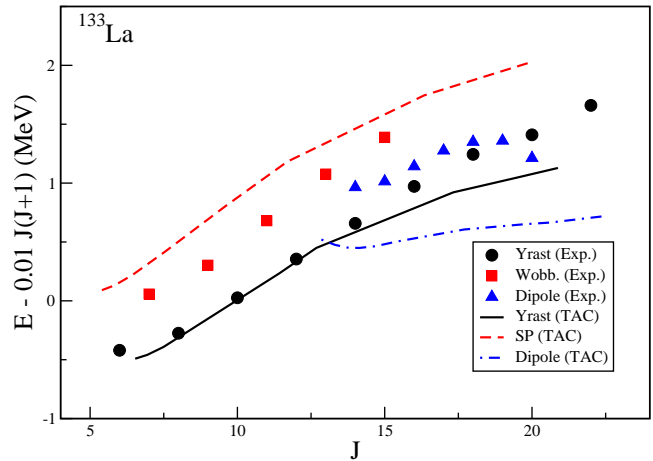}
\caption{\label{fig:6}(color online). Experimental (symbols) and TAC (lines) level energies minus a rotor contribution for
the yrast (Black circle and solid line),
wobbling (Red square), signature partner (Red dashed line) and dipole (Blue triangle and dash-dotted line) bands.}
\end{figure}

Additionally we carried out Tilted Axis Cranking calculations (TAC) \cite{TAC00} using the equilibrium deformation parameters
$\varepsilon=0.16$, $\gamma=26^\circ$ and the pair fields
$\Delta_p=0.7$ MeV, $\Delta_n=1.0$ MeV. Rotation about the short axis turned out to be stable for the two signatures of the $h_{11/2}$ 
one-quasiproton configuration.   
The distance of the unfavored $\alpha=1/2$ routhian to the favored $\alpha=-1/2$ routhian is included in Fig. \ref{fig:3}. The signature
splitting is substantially larger for 
$Z=57$ than $Z=59$ because of the lower location in the $h_{11/2}$ shell. In case of $^{135}$Pr, the TAC calculations place it somewhat
higher than the wobbling band,
which is consistent the experimental localization very close to it. 
 In case of $^{133}$La, TAC predicts it substantially higher, which explains the fact that we could not identify it in the present experiment.
 
 As for $^{135}$Pr, we assign the configuration [$\pi\mathrm{h}_{11/2}\nu\mathrm{h}_{11/2}^2$]  to the dipole band. We carried out TAC calculations
 occupying  the lowest pair of the  $\nu\mathrm{h}_{11/2}$ routhians .  The equilibrium deformation of $\varepsilon=0.16$, $\gamma=40^\circ$ was found.
 The  tilt angle $\vartheta$ with respect to the long axis stays around 35$^\circ$. Below $J=13\hbar$, the second tilt angle is $\varphi=0$, i.e. 
 the rotational axis lies in the short-long plane. Above $J=13\hbar$, the tilt angle $\varphi$ rapidly increases, reaching $56^\circ$ at  $J=22.4\hbar$,
 which suggests the 
 existence of a chiral partner of the dipole band. As seen in Fig. \ref{fig:6}, the TAC calculations well reproduce the position of the dipole band relative
 to the yrast band.
 The TAC calculations also give large values of $B(M1, I\rightarrow I-1)$, which decrease from $3.8$ to $2.0\mu_N^2$ over the shown spin range. Together with the 
 very small TAC
   inband values of $B(E2, I\rightarrow I-2)_{in}<0.02(eb)^2 $ the theory suggests that the band has the character of magnetic rotation. 
  This seems  consistent  with the large experimental ratios 
 $(1.6\pm1.2,~11.6\pm1.2,~16.9\pm1.7,~3.6\pm0.4,~ 6.8\pm0.7)(\mu_N/eb)^2$ observed for the transitions originating respectively from the 
 $31/2^-,~ 33/2^-,~ 35/2^-,~ 37/2^-$ states.  
  The TAC ratios  are larger; however, one expects that the mean field approach underestimates the  $B(E2, I\rightarrow I-2)_{in}$ in the case of small 
  static deformation.

In summary, the nature of the wobbling mode in $^{133}$La has been 
investigated. 
Mixing ratios of interband 
transitions obtained from the angular distribution and DCO-polarization measurements indicate
their strong $E2$ nature.
Surprisingly,  $^{133}$La shows longitudinal
wobbling as the wobbling frequency is increasing with spin. This is the first observation of {\it longitudinal} wobbling band in nuclei.
The quasiparticle plus triaxial rotor model (with and without harmonic frozen approximation), was able to reproduce the experimental energies and
transition rates, and clearly point to a
transition from transverse wobbling in $^{135}$Pr to longitudinal wobbling 
in $^{133}$La.
The change from  transverse wobbling to longitudinal wobbling  
is understood as follows:  In $^{135}$Pr only the $h_{11/2}$ quasiproton is aligned with the short axis of the triaxial
density distribution and the medium axis has the largest MoI. This arrangement gives transverse wobbling. 
In $^{133}$La  an additional  pair of positive parity (dg)  quasiprotons aligns early and gradually with the short axis. The additional
alignment increases the effective MoI of the short axis, which becomes larger than the MoI of the medium axis. 
This arrangement gives longitudinal wobbling.   In $^{135}$Pr the early (dg) neutron alignment occurs 
later as part of the sharp band crossing, above which the longitudinal wobbling begins to appear~\cite{ma15}.

We acknowledge the TIFR-BARC Pelletron Linac Facility for
providing good quality beam. The help and cooperation of the INGA collaboration in setting 
up the array is also acknowledged. We thank the TIFR central workshop for the 
fabrication of various mechanical components for the experiment. This work has been supported in part by the Department of
Science and Technology, Government of India (No. IR/S2/PF-03/2003-II),
the U.S. National Foundation (Grant No. PHY-1419765), and the US Department of Energy 
(Grant DE-FG02-95ER40934).  \\\\

\end{document}